\documentclass[a4paper,american,aps,citeautoscript,floatfix,preprintnumbers,prl,showpacs,superscriptaddress,%
longbibliography,
twocolumn]{revtex4-1}

\usepackage{amsfonts,amsmath,amssymb}
\usepackage{graphicx}
\usepackage[T1]{fontenc}
\usepackage[utf8]{inputenc}
\usepackage{microtype}
\usepackage{pxfonts,tgpagella}
\usepackage{subfigure}
\usepackage{textcomp}
\usepackage[textsize=footnotesize]{todonotes}
\usepackage{xspace}
\usepackage{hyperref, hypernat}
\usepackage[todos]{CMImacros}

\newlength{\figwidth}
\newcommand{\HHO}{\ensuremath{\text{H}_2\text{O}}\xspace}
\newcommand{\HHHOp}{\ensuremath{\text{H}_3\text{O}^+}\xspace}
\newcommand{\ind}{\ensuremath{\text{C}_8\text{H}_7\text{N}}\xspace}
\newcommand{\indw}{\ensuremath{\ind(\HHO)}\xspace}
\newcommand{\indmH}{\ensuremath{\text{C}_8\text{H}_6\text{N}}\xspace}

\newcommand{\cfeldesy}{\affiliation{Center for Free-Electron Laser Science, Deutsches
      Elektronen-Synchrotron DESY, Notkestrasse 85, 22607 Hamburg, Germany}}%
\newcommand{\uhhcui}{\affiliation{The Hamburg Center for Ultrafast Imaging, Universität Hamburg,
      Luruper Chaussee 149, 22761 Hamburg, Germany}}%
\newcommand{\uhhchem}{\affiliation{Department of Chemistry, Universität Hamburg,
      Martin-Luther-King-Platz 6, 20146 Hamburg, Germany}}%
\newcommand{\uhhphys}{\affiliation{Department of Physics, Universität Hamburg, Luruper Chaussee 149,
      22761 Hamburg, Germany}}%
\newcommand{\jkemail}{\email{jochen.kuepper@cfel.de}}%
\newcommand{\cmiweb}{\homepage{https://www.controlled-molecule-imaging.org}}%

\begin{document}
\title{Strong laser alignment of solvent-solute aggregates in the gas-phase}%
\author{\mbox{Sebastian Trippel}}\cfeldesy\uhhcui%
\author{\mbox{Joss Wiese}}\cfeldesy\uhhchem%
\author{\mbox{Terry Mullins}}\cfeldesy%
\author{\mbox{Jochen~Küpper}}\jkemail\cmiweb\cfeldesy\uhhcui\uhhchem\uhhphys%
\date{\today}%
\begin{abstract}\noindent%
   Strong quasi-adiabatic laser alignment of the indole-water-dimer clusters, an amino-acid
   chromophore bound to a single water molecule through a hydrogen bond, was experimentally
   realized. The alignment was visualized through ion and electron imaging following strong-field
   ionization. Molecular-frame photoelectron angular distributions showed a clear suppression of
   electron yield in the plane of the ionizing laser’s polarization, which was analyzed as strong
   alignment of the molecular cluster with $\cost\ge0.9$.
\end{abstract}
\maketitle
\noindent

\section{Introduction}
\label{sec:introduction}
The properties of atoms and molecules are strongly dependent on their environment. The photochemical
pathways and functions of biological chromophores are, for instance, strongly affected by
hydrogen-bond interactions with the surrounding protein and solvent
environment~\cite{Tatischeff:CPL54:394, Song:IRPC32:589, Horke:PRL117:163002}. Hydrogen bonds in
general are of universal importance in chemistry and biochemistry and it is of great interest to
bridge the gap between single isolated molecules and molecules in solvation.

Indole is the chromophore of the amino acid tryptophan and indole-water corresponds to a model
system of a chromophore ``solvated'' by a single water molecule. Indole's intrinsic properties have
been widely studied~\cite{Berden:JCP103:9596, Kuepper:PCCP12:4980, Livingstone:JCP135:194307}. Also
the influence of water solvation on indole has been discussed extensively, since it has a strong
influence on its electronic properties~\cite{Sobolewski:CPL329:130}. This includes the energetic of
the lowest electronically excited states, which interchange their order with the addition of water
and other polar molecules~\cite{Lami:JCP84:597}. Its emission properties are regularly used in
fluorescence studies of proteins, where spectral shifts are directly related to the chromophores’
environment~\cite{Vivian:BiophysJ80:2093}. The indole-water dimer has a well-defined
structure~\cite{Korter:JPCA102:7211, Blanco:JCP119:880, Kang:JCP122:174301}, in which water is
hydrogen bonded to the N--H moiety of the pyrrole-unit. The indole-water binding energy was
determined to 0.2~eV~\cite{Mons:JPCA103:9958}.

The rotational motion of molecules can be controlled by the torque exerted on their dipole moment
that is induced by the interaction of the molecule's polarizability with external electric fields.
Intense laser light, providing electric-ac-field intensities on the order of 1~TW/cm$^2$ and above,
can strongly align, and in consequence spatially fix, one or two molecular axes to the laboratory
frame~\cite{Stapelfeldt:RMP75:543}. These aligned molecules serve as ideal samples, \eg, to image
the structure and dynamics of complex molecules directly in the molecular frame. This includes
time-resolved dynamics studies using molecular-frame photoelectron angular distributions
(MFPADs)~\cite{Hansen:PRL106:073001} or photoelectron holography~\cite{Krasniqi:PRA81:033411,
   Landers:PRL87:013002}. In addition, as most chemical reactivity depends on the relative
orientation of the reactants, aligned molecules are well suited to study steric effects in chemical
reactions. Further applications of aligned molecules range from attosecond-light-pulse and
high-harmonic generation~\cite{Velotta:PRL87:183901} over ultrafast electron or x-ray
diffraction~\cite{Filsinger:PCCP13:2076, Hensley:PRL109:133202, Kuepper:PRL112:083002,
   Yang:PRL117:153002} to quantum information processing~\cite{Shapiro:PRA67:013406}. Time-resolved
diffractive-imaging experiments are of utmost interest for the recording of so-called molecular
movies, which is also especially interesting for large and complex molecules. The contrast in these
experiments can be greatly increased if the molecules are fixed in space, \eg, strongly aligned,
typically when $\cost>0.9$~\cite{Filsinger:PCCP13:2076, Barty:ARPC64:415}. Strong molecular
alignment in combination with tomographic approaches to observe three-dimensional diffraction
volumes has the potential to retrieve the bond angles of complex molecules~\cite{Wolf:OptComm1:153}.
The structural dynamics of the indole-water system could be extracted from ultrafast electron or
x-ray-diffraction experiments~\cite{Park:ACIE47:9496, Kuepper:PRL112:083002} on the controlled
system.

Strong alignment was achieved for linear, symmetric top, and asymmetric top molecules in the
adiabatic~\cite{Larsen:JCP111:7774}, intermediate~\cite{Trippel:MP111:1738, Trippel:PRA89:051401R},
and impulsive regimes~\cite{RoscaPruna:PRL87:153902}\footnote{An adiabatic response of the system to
   the laser field is provided if the time scales of the laser pulse are longer than the rotational
   period of the molecule. The impulsive regime is achieved if the pulse duration of the laser is
   much shorter than the rotational period of the molecule.}. Especially the combination with
rotational-state selection~\cite{Filsinger:JCP131:064309, Chang:IRPC34:557} has improved the
achievable control dramatically, the recorded degree of alignment reaching values as high as
$\cost>0.97$~\cite{Holmegaard:PRL102:023001}. For the imaging of complex chemical dynamics, it is
highly desirable to extend the range of directionally-controlled molecular systems to molecular
aggregates with weak intermolecular interactions, such as hydrogen bonded solvent-solute clusters.
These systems also serve as tests for the possibilities to laser-align large biomolecules without
deterioration of the secondary structure~\cite{Spence:PRL92:198102, Barty:ARPC64:415}. Besides the
generation of weak rotational coherences, corresponding to very weak alignment, in spectroscopic
investigations of hydrogen bonded clusters and even very weakly bound molecule-rare-gas
clusters~\cite{Connell:JCP94:4668, Riehn:CP283:297, Blanco:JCP119:880, Galinis:PRL113:043004}, so
far no alignment for such systems has been reported.

Here, we report strong three-dimensional laser alignment of the prototypical indole-water dimer
cluster. This enables for the molecular-frame imaging (\emph{vide supra}) of such complex, floppy,
composite molecular systems. More specifically, we present experimental results concerning the
long-pulse, quasi-adiabatic alignment of the spatially-separated indole-water cluster and its
strong-field-ionization (SFI) molecular-frame photoelectron angular distributions (MFPADs).

\section{Experimental setup}
\label{sec:experimental-setup}
The experimental setup was described elsewhere~\cite{Trippel:MP111:1738}. In brief, a pulsed
molecular beam was provided by expanding a few millibar of indole and a trace of water seeded in
80~bar of helium through an Even-Lavie valve~\cite{Even:JCP112:8068} operated at a temperature of
110~$^{\circ}\mathrm{C}$ and at a repetition rate of 100~Hz, limited by pumping speed. Due to
three-body collisions in the early phase of the expansion $(\text{indole})_m(\HHO)_n$ clusters were
formed. Using the electrostatic deflector, the molecular beam was dispersed according to the
species' and quantum states' dipole-moment-to-mass ratio~\cite{Chang:IRPC34:557,
   Filsinger:JCP131:064309, Trippel:PRA86:033202}.

The indole-water clusters were 3D aligned by a moderately intense
($\Icontrol\approx10^{11}~\text{W/cm}^2$), 485~ps long laser pulse inside a velocity map imaging
(VMI) spectrometer. The rise time in intensity (10--90~\%) was 100~ps~\cite{Trippel:PRA89:051401R}.
The laser pulse was elliptically polarized with an aspect ratio of $3:1$ in intensity, with the
major axis along the laboratory $Z$ axis. The angular confinement was probed through strong-field
multiple ionization by a linearly polarized, 30~fs laser pulse
($\Iprobe=3\cdot10^{14}~\text{W/cm}^2$), resulting in Coulomb explosion of the cluster. The
resulting ions were velocity mapped onto a position sensitive detector. The alignment and probe
laser pulses were provided by an amplified femtosecond laser system~\cite{Trippel:MP111:1738}. The
probe pulses had pulse energies of 200~\uJ and a beam waist of \mbox{$\omega_0=24~\um$}. The
alignment laser pulses had energies controlled between 0 and 7~\mJ and a beam waist of
\mbox{$\omega_0=30~\um$}. The alignment laser itself did not ionize the molecules.

For the MFPAD experiments, a circularly-polarized 30~fs probe pulse centered at 800~nm with a peak
intensity of $\Iprobe=1.3\cdot10^{14}~\text{W/cm}^2$ was used, which places the ionization process
close to the tunneling regime with a Keldysh parameter of $\gamma = 0.7$. The circular polarization
ensures that no recollisions of the electrons with the parent ions occur. Under these conditions,
the cluster undergoes only single ionization, as confirmed by ion time-of-flight mass spectrometry.

\section{Results}
\label{sec:results}
\begin{figure}
   \centering
   \includegraphics[width=\linewidth]{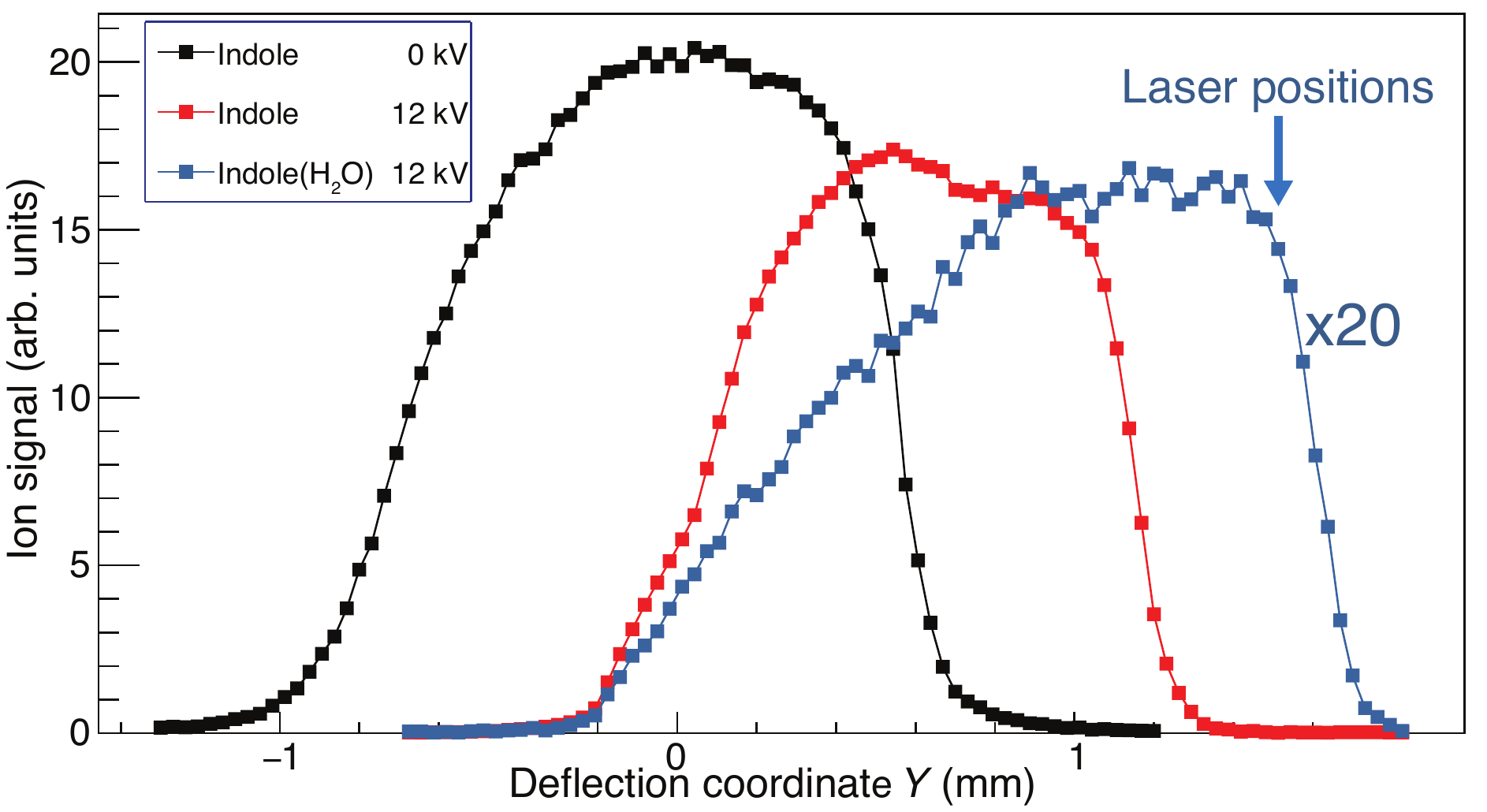}
   \caption{(Color online): Molecular-beam and deflection profiles of indole and indole-water.}
   \label{fig:deflection}
\end{figure}
\autoref{fig:deflection} shows the measured horizontally integrated column density profiles of the
undeflected and deflected molecular beam as a function of the vertical laser position, for further
details see reference~\onlinecite{Chang:IRPC34:557}. The profiles were measured \emph{via}
strong-field ionization of the clusters and molecules in a velocity map imaging spectrometer. The
individual profiles were obtained by gating the detector on the specific parent ion masses. The
undeflected molecular beam column density profile, which has the same shape for all species, is
shown as a black line, whereas the deflected molecular beam column density profiles for indole and
indole-water are plotted in red and blue, respectively. The indole-water profile is multiplied by a
factor of 20 with respect to the undeflected and deflected indole signal, which reflects the
relative amount of this dimer formed in the supersonic expansion. A clear separation of indole and
indole-water clusters was observed~\cite{Trippel:PRA86:033202}. From the various species present in
the beam the indole-water cluster possess the largest dipole moment
($\mu=4.4$~D)~\cite{Kang:JCP122:174301} and the largest dipole-moment-to-mass ratio, and, therefore,
was deflected significantly more than the indole ($\mu=1.96$~D) and water ($\mu=1.86$~D)
monomers~\cite{Horke:ACIE53:11965, Trippel:PRA86:033202}. Similarly, higher order clusters
(indole)$_m(\HHO)_n$ ($m>1\vee{}n>1$) were deflected significantly less. Thus, a pure sample of
indole-water is obtained and the blue arrow marks the vertical position in the molecular beam where
the alignment and MFPAD measurements were performed.

\begin{figure}
   \centering%
   \includegraphics[width=\linewidth]{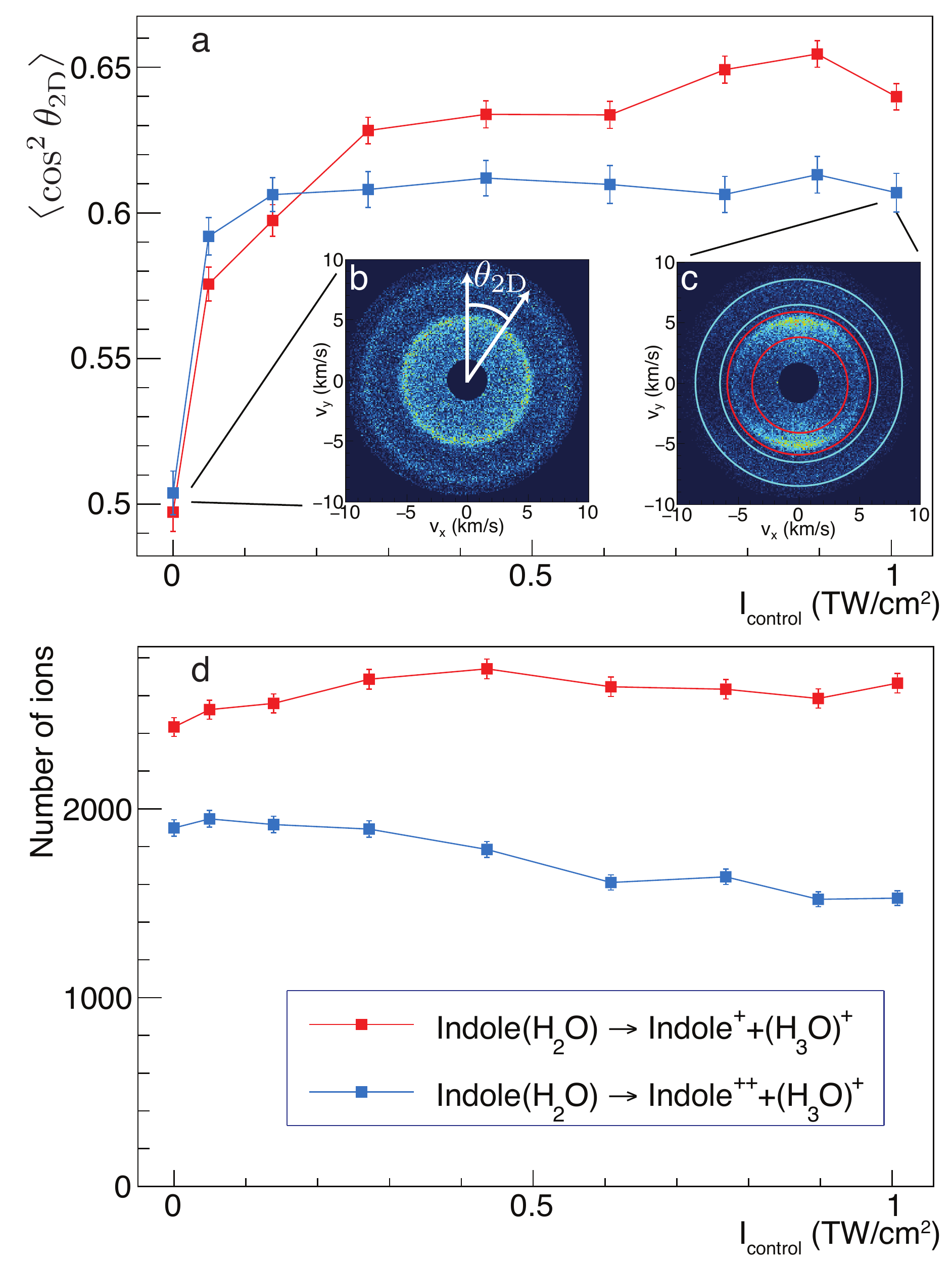}
   \caption{(Color online): a) Observed degree of alignment of \HHHOp as a function of the control
      laser peak intensity for the fragmentation channels with singly (red) and doubly (cyan)
      charged indole fragments. The angle $\thetatwoD$ is defined in b. Velocity maps b) without
      alignment laser and c) with alignment laser at a peak intensity of 1~TW/cm$^2$. Pairs of
      circles indicate the corresponding velocity cuts for the two channels. d) Number of ions for
      the singly (red) and doubly (cyan) indole-charge separation channel inside the velocity
      selected areas.}
   \label{fig:powerdependence}
\end{figure}
The 2D momentum image for \HHHOp ions for the alignment-field-free case is shown in
\subautoref{fig:powerdependence}{b}. Clearly visible are two, a slow and a fast, ionic channels,
assigned to the two Coulomb explosion channels $\indw\longrightarrow\indmH^{+}+\HHHOp$ and
$\indw\longrightarrow\indmH^{++}+\HHHOp$. The central part of the image was cut due to the presence
of background signal in this region. The observed distribution is circularly symmetric as expected
for the case of an isotropic sample and the probe-laser polarization linear and perpendicular to the
detector plane. \subautoref{fig:powerdependence}{c} shows the corresponding ion distribution when
the molecules were 3D aligned by the elliptically polarized control laser with its major axis
perpendicular to the detector plane. The pairs of red and cyan circles indicate the velocity ranges
that were used to extract the degree of alignment for the two channels. This \HHHOp distribution was
fairly directional and provided direct evidence of the alignment of the indole-water clusters at the
time of ionization. The mean of the momentum distribution of the $\HHHOp$ fragments was located in
the molecular plane of the indole moiety due to the planarity of the indole-water cluster. Measuring
its transverse momentum distribution, therefore, directly linked the molecular plane of the indole
moiety to the laboratory frame.

\subautoref{fig:powerdependence}{a} shows the experimentally derived degree of alignment
\cost~\footnote{The two-dimensional degree of alignment is defined as
   $\cost=\int_{0}^{\pi}\int_{0}^{r_{max}}\!\cos^2\!\left(\theta_\text{2D}\right)
   f(\thetatwoD,r_{2D}) \, dr_{2D}d\thetatwoD$. $f(\thetatwoD,r_{2D})$. The degree of alignment is
   extracted from the raw data by accepting only ion events in the integrals with transverse
   velocities within the specific velocity regions depicted in \subautoref{fig:powerdependence}{a}.}
as a function of the peak intensity of the alignment pulse for the slow,
$\indw\longrightarrow\indmH^{+}+\HHHOp$, and fast, $\indw\longrightarrow\indmH^{++}+\HHHOp$,
channels in red and cyan, respectively. Without alignment laser a degree of alignment $\cost=0.5$
was observed. With increasing laser intensity, the degree of alignment increased. It saturated for
both channels at relatively moderate intensities.

\subautoref{fig:powerdependence}{d} shows the integrated number of \HHHOp ions for the two channels
within the corresponding velocity cuts as a function of the alignment laser peak intensity. For the
singly-charged-indole fragmentation channel the number of ions increased with the peak laser
intensity is observed. In contrast, it decreased for the doubly-charged-indole fragmentation
channel.

\begin{figure}
   \centering
   \includegraphics[width=\linewidth]{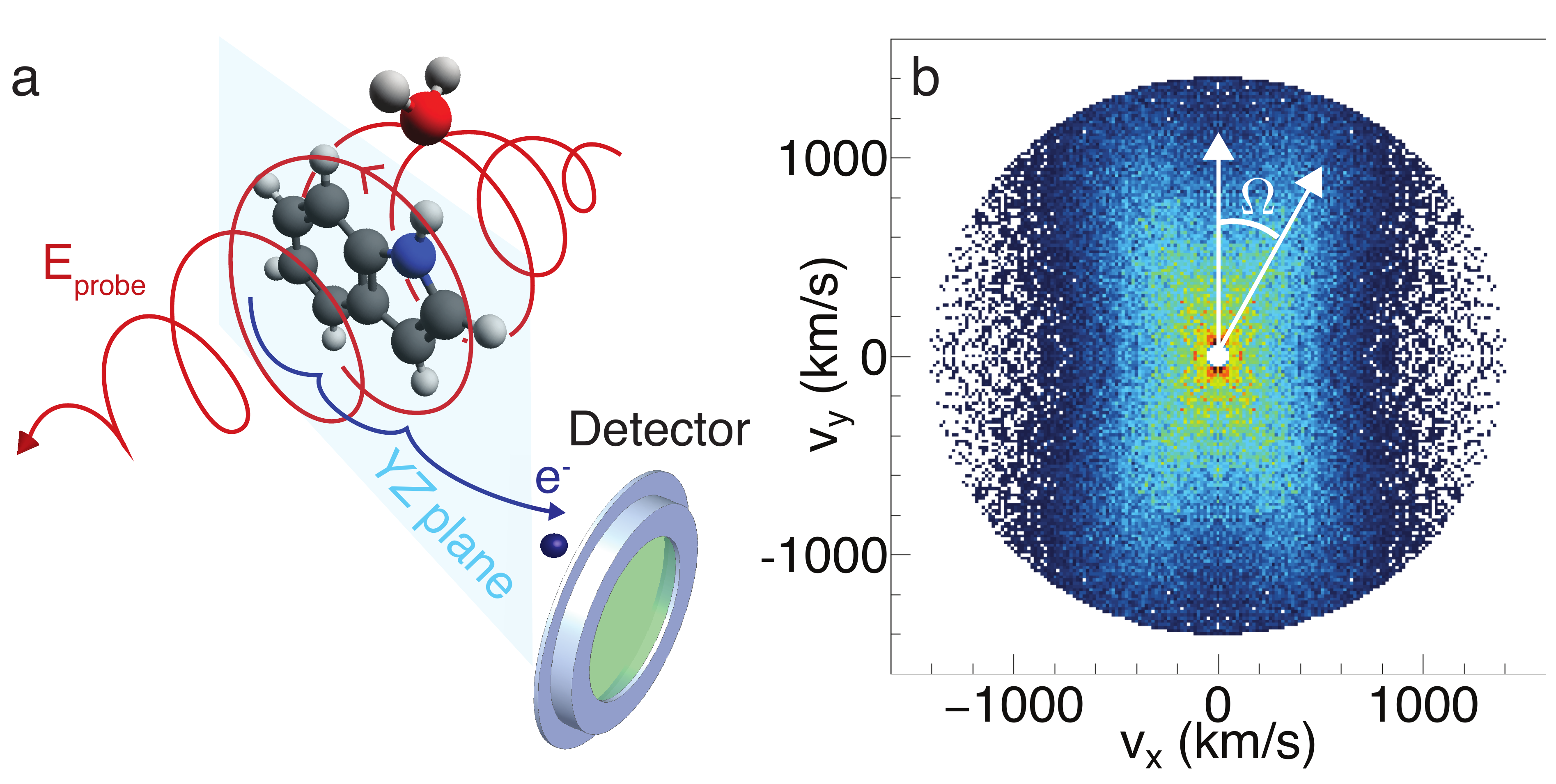}
   \caption{(Color online): a) Schematic of the experimental setup showing the indole-water cluster
      aligned to the $YZ$ plane (cyan). The circularly polarized probe laser is indicated by the red
      spiral. b) Experimentally observed MFPAD of indole-water.}
   \label{fig:mfpads}
\end{figure}
\autoref{fig:mfpads}~a shows a sketch of the MFPAD measurements. Indole-water clusters were aligned
to the $YZ$-plane) as described above and singly ionized by a circularly polarized probe laser,
indicated by the red spiral. The emitted electrons are accelerated by the combined static electric
field in the spectrometer and the laser field and were projected onto a position sensitive detector.
\subautoref{fig:mfpads}{b} shows the recorded MFPAD of the aligned indole-water clusters. The
electron emission is suppressed along the vertical direction, clearly seen as dips in the
corresponding angular distributions. The maximum in the off-the-nodal-plane angle in the electron
emission~\cite{Holmegaard:NatPhys6:428}, occurs at an angle $\Omega=\degree{22}$ with respect to the
the $YZ$-plane.

\section{Discussion}
\label{sec:discussion}
Indole-water has a longest classical-rotation period of 484~ps. The control laser pulse duration of
485~ps with a rising edge of about 100~ps places the alignment dynamics of the cluster into the
intermediate regime between impulsive and adiabatic alignment~\cite{Trippel:PRA89:051401R}.
Different from observations for rotational-ground-state-selected OCS, no oscillations in the degree
of alignment as a function of the delay between control-laser and probe-laser pulses are observed.
While for every initial rotational state an oscillatory behavior in the degree of alignment should
be induced, these effects are averaged out due to the different phases of these oscillations for
different initial states and the large number of initial rotational states present in the deflected
part of the beam for larger molecules~\cite{Filsinger:JCP131:064309}.

The observed power dependence of the degree of alignment for both channels in
\subautoref{fig:powerdependence}{a} is similar to the typical dependence observed for a cold
molecular beam~\cite{Holmegaard:PRL102:023001}. The degree of alignment increases strongly with
increasing control laser power until it reaches a plateau at relatively low control laser
intensities. It differs from experiments with good indicators for the degree of alignment in the
actual degree of alignment of the plateau, which is given by $\cost=0.64$ and $\cost=0.61$ for the
singly and doubly charged fragmentation channels respectively. For a strongly aligned sample with
good axial recoil a limit of $\cost>0.95$ is typically observed. On the other hand, a slow increase
in the degree of alignment is usually observed as a function of the peak intensity of the alignment
laser for a warm molecular beam~\cite{Kumarappan:JCP125:194309}. In that case, the degree of
alignment does not reach the plateau region for laser intensities in the order of 1~TW/cm$^2$.
Therefore, the saturation behavior observed here is attributed to strong alignment of cold molecules
and the poor representation of this alignment is attributed to non-axial recoil in the molecular
frame, likely due to complex, possibly slow, chemical dynamics of the formation of
\HHHOp.

Knowledge of the in-plane momentum distribution of the $\HHHOp$ fragments in the molecular frame
would, in addition, allow to determine the three-dimensional degree of alignment and orientation of
the cluster by measuring the three-dimensional momentum vector. However, this is not trivial as this
dynamics in the molecular frame is unknown and an analysis through complex quantum-chemical
computations is beyond the scope of this publication. Moreover, as the measured degree of alignment
was saturated at a degree of alignment of $\cost=0.64$, the directionality of this probe process is
also not expected to be highly directional inside the molecular plane of the indole moiety. On the
other hand, three-dimensional alignment is routinely obtained for a control-laser-pulse ellipticity
of 3:1 if the one-dimensional alignment is strong~\cite{Larsen:PRL85:2470, Nevo:PCCP11:9912}.

The strong degree of planar alignment is supported by the recorded MFPADs, which contain information
on the relative dipole moments, polarizabilities, molecular orbitals involved in the ionization, and
the laser-field-dependent ionization potentials~\cite{Holmegaard:NatPhys6:428,
   Kjeldsen:PRA71:013418}. The local minimum of electrons along the vertical direction results from
suppression of electron emission in the nodal plane of the highest-occupied molecular orbitals (HOMO
and HOMO-1)~\cite{Holmegaard:NatPhys6:428}. Instead, due to the symmetry of these $\pi$ orbitals,
the emission occurs off the nodal plane in the molecular frame. The suppression manifests itself by
a dip along the vertical axis of the measured MFPAD since the nodal plane of the molecule is
strongly fixed to the laboratory frame ($YZ$ plane). This feature appears typically if the molecules
are strongly aligned with a degree of alignment $\cost\ge0.9$. The measured opening angle
$\Omega=\degree{22}\pm\degree{2}$ is in good agreement with the theoretical value of
$\Omega_{\text{theo}}=\degree{20}$ obtained within the extended strong-field ionization
model~\cite{Holmegaard:NatPhys6:428} using a computed ionization potential of the cluster of
$I_\text{p}=7.71$~eV ($\text{MP2/6-311G}^{**}$). Rotating the major axis of the ellipse of the
control-laser polarization with respect to the detector allowed to retrieve the three-dimensional
MFPAD of indole-water by tomographic reconstruction methods, similar to previous work for
napthalene~\cite{Maurer:PRL109:123001}. Preliminary analysis of this data further confirms this
argument.

The limit of the measured degree of alignment for the slow channel is higher than for the fast
channel. This is surprising as the velocity for the doubly charged indole channel is faster and thus
better axial recoil is expected. We attribute this behavior to a change in the branching ratios of both
channels as a function of the peak intensity of the alignment laser. The faster channel is
suppressed in the strong field compared to the slow channel, see
\subautoref{fig:powerdependence}{d}. This signal decrease might be attributed to the symmetry plane
of the involved orbitals, which leads to a reduced ionization probability when the molecular frame
is fixed to the laboratory frame (\emph{vide supra}). For the most-strongly-aligned molecules,
situated and probed at the peak of the alignment laser intensity distribution, multiple ionization
is suppressed and they are preferentially observed in the slow channel. This leads to a faster
saturation effect in the measured degree of alignment for the doubly charged indole channel as a
function of the laser intensity. The degree of alignment for the singly charged channel is still
slightly increasing due to the increase in the degree of alignment.

One could speculate that the cluster might be destroyed by the control laser itself. A possible
neutral destruction mechanism would lead to a reduction of the \HHHOp Coulomb explosion channels, as
the pulse duration of the control laser is relatively long compared to a typical dissociation
timescale. However, the constant number of \HHHOp fragments for the whole range of control-laser
intensities indicates that the clusters are not destroyed by the alignment laser. This is consistent
with a small probability expected for a transition of a bound state to the continuum wavefunction,
which leads to a highly inefficient direct fragmentation of the cluster in the electronic ground
state by the absorption of a single photon at the wavelength of the alignment laser. In addition,
while electronically excited states of indole-water could break the cluster by a charge transfer in
the $A''(\pi\sigma^*)$ state~\cite{Sobolewski:CPL329:130}, the three-photon-transition probability to
populate these states by the alignment laser is very small~\cite{Lippert:CPL376:40}. Moreover, the
laser-induced Stark effect, which is on the order of a few meV for the alignment laser intensity,
is not relevant for breaking the intermolecular hydrogen bond in the cluster, which is bound by
0.2~eV.

\section{Conclusions}
\label{sec:conclusions}
The hydrogen-bound indole-water clusters were strongly quasi-adiabatically aligned by moderately
strong elliptically-polarized laser pulses. Alignment was observed in velocity-map images of \HHHOp
fragments, albeit only a weak degree of alignment, $\cost=0.64$, could be deduced, which is
attributed to non-axial recoil. Much stronger alignment, $\cost\ge0.9$, was deduced from a clearly
visible nodal plane in the molecular-frame photoelectron angular distributions. This strong control
opens up possibilities to study the photo-induced ultrafast chemical dynamics around the
intermolecular hydrogen bond~\cite{Conde:CP530:25} through the imaging of half-collisions in the
molecular frame, which could provide unprecedented details on these intermolecular interactions.

The current results highlight that the determination of molecular axes through fragment recoil leads
not always to the correct determination of the degree of alignment in the laboratory frame.
Reconstructing the MFPADs from the recoil frame distribution is, therefore, not possible for such
systems. Our findings underline the advantage of active laser alignment, in which the laser
polarization fixes the molecular frame to the principal axes of polarizability. However, a possible
influence of the alignment laser on the investigated dynamics has to be considered.
Field-free-alignment with shaped laser pulses could provide the best solution, but in that approach
it is difficult to achieve strong alignment for asymmetric tops, and we are currently exploring
novel schemes for complex molecules.

\section{Acknowledgements}
Besides DESY, this work has been supported by the \emph{Deutsche Forschungsgemeinschaft} (DFG)
through the excellence cluster ``The Hamburg Center for Ultrafast Imaging -- Structure, Dynamics and
Control of Matter at the Atomic Scale'' (CUI, EXC1074) and the priority program 1840 ``Quantum
Dynamics in Tailored Intense Fields'' (QUTIF, KU1527/3), by the European Research Council through
the Consolidator Grant COMOTION (ERC-Küpper-614507), and by the Helmholtz Association ``Initiative
and Networking Fund''.

\bibliography{string,cmi}
\end{document}